\documentclass[11pt,a4paper,english,nofootinbib,,superscriptaddress]{revtex4-2}

\usepackage{lmodern}

\usepackage[T1]{fontenc}
\usepackage[latin9]{inputenc}
\usepackage{babel}
\usepackage{color}

\usepackage{amsmath}
\usepackage{graphicx}
\usepackage{amssymb}
\usepackage{esint}
\usepackage[unicode=true, pdfusetitle,
 bookmarks=true,bookmarksnumbered=false,bookmarksopen=false,
 breaklinks=false,pdfborder={0 0 1},backref=false,colorlinks=false]
 {hyperref}
\usepackage{amsmath}
\usepackage{amssymb}
\def\b{\begin{equation}}
\def\e{\end{equation}}

\def\1{\mbox{I\hspace{-.15em}1}}

\def\R{{\rm I\hspace{-.15em}R}}

\def\C{\hspace{3pt}{\rm l\hspace{-.47em}C}}

\def\b{\begin{equation}}
\def\e{\end{equation}}
\def\bee{\begin{enumerate}}
\def\eee{\end{enumerate}}

\def\R{{\rm I\hspace{-.15em}R}}

\def\C{\hspace{3pt}{\rm l\hspace{-.47em}C}}

\def\bd{\begin{displaystyle}}
\def\ed{\end{displaystyle}}
\def\ba{\begin{array}}
\def\ea{\end{array}}

\def\bee{\begin{enumerate}}
\def\eee{\end{enumerate}}

\def\bes{\begin{eqnarray*}}
\def\ees{\end{eqnarray*}}
\def\be{\begin{eqnarray}}
\def\ee{\end{eqnarray}}

\makeatletter
\@ifundefined{textcolor}{}
{%
 \definecolor{BLACK}{gray}{0}
 \definecolor{WHITE}{gray}{1}
 \definecolor{RED}{rgb}{1,0,0}
 \definecolor{GREEN}{rgb}{0,1,0}
 \definecolor{BLUE}{rgb}{0,0,1}
 \definecolor{CYAN}{cmyk}{1,0,0,0}
 \definecolor{MAGENTA}{cmyk}{0,1,0,0}
 \definecolor{YELLOW}{cmyk}{0,0,1,0}
 }

\usepackage{latexsym}\usepackage{bm}

\makeatother

\begin{document}
\title{Quantum states and quantum computing}
\author{Mohammad Vahid Takook,}
\affiliation{APC, UMR 7164, Universit\'e Paris Cit\'e, 75205 Paris, France}
\email{takook@razi.ac.ir}
\author{Ali Mohammad-Djafari}
\affiliation{Laboratoire des Signaux Systemes, Centrale Supelec,  Universite Paris Saclay and CNRS, France}


\begin{abstract}
In classical theory, the physical systems are elucidated through the concepts of particles and waves, which aim to describe the reality of the physical system with certainty. In this framework, particles are mathematically represented by position vectors as functions of time, $\vec{x}(t)$, while waves are modeled by tensor fields in space-time, $\Phi(t, \vec{x})$. These functions are embedded in, and evolve within space-time. All information about the physical system are coded in these mathematical functions, upon which the classical technologies are developed. In contrast, quantum theory models the physical system using a quantum state $\vert \alpha ,t\rangle$, situated in an evolving within Hilbert space, portraying the system's reality with inherent uncertainty. Despite the probabilistic nature of reality observation, the quantum state $\vert \alpha ,t\rangle$ can be precisely determined due to the unitary principle, provided we know the initial state. Therefore, it can serve as a foundation for developing quantum technologies, which we call quantum state-tronics similar to electronics. This discussion focuses on quantum computation, given its expansive scope. One of the paramount challenges in quantum computing is the scarcity of individuals equipped with the requisite knowledge of quantum field theory and the training necessary for this field. This article aims to elucidate the fundamental concepts of quantum field theory and their interconnections with quantum computing, striving to simplify them for those engaged in quantum computing.\end{abstract}
\maketitle




\section{Introduction}

In classical theory, the physical systems are explained by the concepts of particles and waves, aiming to explain the reality of the physical system with certainty. In these cases, the particle is mathematically modeled by the position vector as a function of time or the trajectory of the particle in space-time,  $\vec x(t)$, and the wave is modeled by a tensor (or spinor) field on space-time,  $\Phi(t,\vec x)$. They are immersed in and evolve within space-time, and all information concerning the physical system is encoded in these mathematical functions and the technology is constructed over them. The space-time is modeled by a rank-$2$ symmetric tensor field $g_{\mu\nu}(t,\vec x)$ in the general relativity context \cite{swgr}. It can be visualized as a $4$-dimensional hyper-surface or manifold $M$. Therefore, the primitive concepts are spacetime, particles, and waves. The equation of motion of these functions is obtained from the principles and laws of classical physics, for instance, Newtonian laws, special and general relativity principles, laws of thermodynamics, Gauss's law, Ampere's circuital law, etc. By solving the equations of motion and imposing the initial or boundary condition these functions can be obtained deterministically. The observation of the physical system in the spacetime can obtain the initial conditions.The classical technology is constructed by using these functions, $\vec x(t)$ or $\Phi(t,\vec x)$, which will be briefly recalled in Section \ref{Classical notions}.

Quantum technology has two different periods of progress which are based on the first quantization (quantum mechanics) and the second quantization (quantum field theory). In the first quantization, the mathematical model of the particle such as an electron, $\vec x(t)$, is replaced by a spinor field $\psi(t,\vec x)$. The quantum behavior of the particle is probabilistic, while its associated spinor field $\psi$ is classical and deterministic. The first quantum industrial revolution began with observing electrons' wave properties, which came from quantizing a particle, {\it i.e} a wave function describes a particle, $\psi(t,\vec x)$, which is a classical field. However, the observed properties of the particle (such as position, velocity, energy, momentum, etc) are explained with a probability density function in space-time $\vert\psi(t,\vec x)\vert^2$. The first quantization leads to the first quantum technology revolution. Technologies generated by the first quantization are electron microscope, nuclear power, Laser, transistors and semiconductor devices, and other devices such as MRI devices. The first quantization can be considered as a classical field theory \cite{takook2} and the physical system in this domain of energy can be simulated over a classical computer \cite{fey}.

Although the first quantization or classical field theory successfully explains the structure of atoms and predicts many fundamental particles, it suffers from many practical and philosophical problems, such as the creation and annihilation of particles and particle-wave duality. These problems are pretty solved in second quantization or quantum field theory (QFT). In QFT, the physical systems are modeled with a quantum state $\vert\alpha \rangle$, which is immersed in the Hilbert space $\mathcal{H}$. Therefore the primitive fundamental concepts in QFT  are the quantum state of a physical system and Hilbert space. The mathematical models of particle $\vec x(t)$ and wave $\Phi(t,\vec x)$ are replaced by the quantum state $\vert\alpha \rangle$, and the Hilbert space $\mathcal{H}$ plays a role analogous to that of the spacetime manifolds $M$ and it is a fiber bundle over spacetime $\mathcal{H} \times M$ \cite{tasvu,taqg,ta20}. We briefly recall the QFT notions in Section \ref{QFT notions}. 

The second quantum industrial revolution was begun by observing the single photon or particle properties of the radiation field, and wave properties of the single-photon \cite{aspect0,Aspect}. It is based on quantum state properties. The construction procedure of the quantum technology in this case may be performed in three steps: 
\begin{itemize}
\item The construction of a suitable quantum state $\vert \beta ,t_0\rangle$.
\item Obtaining its time evolution under the influence of the interaction Hamiltonian $H_n$: 
$$\vert\beta,t\rangle=U(t,t_0;H_n)\vert\beta,t_0\rangle\,,$$
where $U$ is the time evolution operator.
\item Detection or observation of the quantum state, $\vert\beta,t\rangle\longrightarrow\vert\gamma\rangle$. 
\end{itemize}
This process will be briefly discussed in Sections \ref{logicgate}. A class of devices actively create, manipulate, and read out quantum states of a system, often using the quantum effects of superposition and entanglement. The second quantum technologies revolution is constructed on the quantum state $\vert\alpha,t\rangle$. This model results in quantum technology such as quantum computing, quantum sensors, quantum cryptography, quantum simulation, quantum metrology, quantum imaging, and high-power fiber laser, etc., for more details and references see \cite{takook2}. The physical system in this domain of energy cannot be simulated over a classical computer and we need a quantum computer for simulation of the physical system, which is described by a quantum state $\vert\alpha \rangle$ \cite{oritz}.

The most challenging aspect of quantum technology arises during interactions that result in the reduction or collapse of the quantum state. This includes the interaction of the quantum state with the environment and its detection. According to the principles of quantum theory, the quantum state of a physical system is defined only for isolated systems. However, the process of performing calculations - which involves inputting information into the system, processing calculations, and finally reading out the information - necessitates interaction with the environment. These interactions introduce errors that must be addressed through a process known as quantum error correction. The errors are the unwanted changes to a quantum state caused by interactions with the environment or during the measurement process, which need to be corrected to maintain the integrity of quantum computations or communications.

\section{Classical notions}\label{Classical notions}

In classical theory, the three fundamental concepts are spacetime, particles (or matter), and tensor (or radiation) fields. In the modern formulation of physical systems, group theory plays a central role. The symmetry group of spacetime and the internal symmetries of physical systems are utilized to extract the fundamental concepts of spacetime, particles, and tensor fields. Additionally, the interactions between them can be formulated from the perspective of local symmetrical groups, or gauge theory \cite{bailse}.


\subsection{Space-time}

The spacetime in the small velocity and very small gravitational field can be modeled by the Galilean relativity. In this case, the space and time are invariant under the Galilean group. Time, space, and matter-radiation are independent of each other. The time can be modeled with the real number $\R$ and space with $\R^3$, which are the Euclidean geometry. Time and space may be represented by $M \equiv \{ \mathbb{R} , \mathbb{R}^3\}$, and presented by $t$ and, $\vec x=(x,y,z)$ respectively.

In Special Relativity, space and time are not independent; instead, they are interwoven into a single continuum known as spacetime. They are independent of matter and radiation. Spacetime is flat, represented as $M = \mathbb{R} \times \mathbb{R}^3$, and is known as Minkowski space and a point in the spacetime is presented by a four-vector $x^\mu\equiv (t, x,y,z)$. The spacetime is invariant under the Poincar\'e group. 

In General Relativity, the matter and radiations are the source of the spacetime and in the presence of the source, the spacetime is curved. It can be visualized as a $4$-dimensional  hyper-surfaces or manifold $M$, which is immersed in a flat $5$-dimensional Minkowski space.  All necessary information about the curved spacetime is encoded in the metric tensor field $g_{\mu\nu}(x^\mu)$, which can be obtained by the solution of the Einstein field equation.

\subsection{Particles}

 An elementary particle has intrinsic properties such as mass $(m)$, electrical charge $(q)$, spin $(s)$, flavor $(f)$ and color $(c)$, where the numerical values can be determined experimentally. The last three properties can be measured at the atomic and subatomic levels. These intrinsic properties are associated with four fundamental interactions in nature and play an important role in technology.
 
 The concept of a particle's mass and spin can be extracted from the Poincar\'e group and the electrical charge, flavor and color from the $U(1)$, $SU(2)$ and $SU(3)$ respectively. Mass is used in mechanical technology and the electric charge play the central role in electronic and it is also the source of the electromagnetic waves. The particle is mathematically modeled by the trajectory of the particle in the spacetime $\vec x(t)$, which contains all necessary information about the physical system of particles.

\subsection{Waves}

The intrinsic properties of a wave include amplitude, phase, frequency, wavelength, and polarization. The wave is mathematically modeled by a tensor (or spinor) field $\Phi(t,\vec x)$. All necessary information about the physical system is encoded in this function, such as the energy, momentum, and angular momentum, which they are carried by the wave.

 The free field equation is derived using the Casimir operators of the symmetrical group of spacetime, allowing the definition of the free Lagrangian density. The Lagrangian interaction density is defined through the principal bundle and its connection in differential geometry (gauge theory).  It can be considered as a fiber in a bundle over spacetime. For the electromagnetic field, it can be locally visualized as a product space $U(1)\times M$. If we have the initial or boundary conditions, the solution of the field equations allows us to calculate $\Phi(t,\vec x)$ deterministically. A particle is localized at a point in spacetime, whereas a wave has a distribution over spacetime.


\subsection{Classical logic gate} \label{clalogicgate}

In classical computation, a physical gate that performs logical operations can be constructed using a variety of electronic components. The most fundamental building blocks include Transistors, Diodes, Capacitors, Resistors, Relays and etc. These components can be combined in various ways to create the fundamental logic gates used in classical computation, such as AND, OR, NOT, NAND, NOR, XOR, and XNOR gates. Each gate type performs a basic logical operation, and by combining these gates in different configurations, you can perform complex computations. This is the basis of digital electronics and classical computing as we know it today. They work by manipulating the electric current to flow to the circuit. For example, Transistors can act as switches or amplifiers and are used to create logic gates. By controlling the flow of current through a semiconductor material, transistors can switch between on and off states, corresponding to the binary values $1$ and $0$, respectively, {\it i.e.} a classical bit.
In the one-dimension, the electric current in a wire with the cross-sectional area $A$, is
$$ I={\frac {\mathrm {d} Q}{\mathrm {d} t} =nAq v= nAq \frac {\mathrm {d}  x(t)}{\mathrm {d} t}\,,}$$
where $n$ is the number of charged particles per unit volume (or charge carrier density), $q$ is the charge on each particle and $v=\frac {\mathrm {d}  x}{\mathrm {d} t}$ is the drift velocity.

For instance, Designing an XOR (Exclusive OR) gate can be approached in various ways depending on the components you wish to use. An XOR gate is a digital logic gate that outputs true, $1$, only when the inputs are different. A simple and commonly used method is with metal-oxide-semiconductor field-effect transistors or bipolar junction transistors as they offer a compact and efficient way to build the gate, which is based on wave properties.

\section{Quantum field theory notions}\label{QFT notions}

In QFT, the physical system is modeled by a quantum state vector $|\Psi(\nu,n)\rangle$, where $\nu$ and $n$ are the set of continuous and discrete quantum numbers respectively. They are labeled the eigenvalue and eigenvector of the set of commutative operator algebras of the physical system and determine the Hilbert space \cite{tagahu}. Although the particle and tensor(-spinor) fields, $\Phi(\vec x,t)$, are immersed in a spacetime manifold $M$, the quantum state vector is immersed in a Hilbert space $ \mathcal{H}$. Therefore the fundamental concepts in QFT are the Hilbert space and the quantum state, which will be briefly recalled in this section.

\subsection{Hilbert space} In quantum theory the observables of the physical system are described by the operators and then the Hilbert space is constructed from the operator algebra of these observables. Utilizing the Lie algebra of the spacetime symmetry group, Poincar\'e group algebra, one can construct the one-particle Hilbert space $ \mathcal{H}^{(1)}$.  The field operator is defined as a map on the Hilbert space. It assumes that it is constructed by the annihilation and creation operators on the Hilbert space \cite{bailse,brmo96}:
\b  \label{ancr}  \Phi=\Phi^-+\,\Phi^+\, ,  \e 
where $\Phi^+$ creates a one-particle state and $\Phi^-$ annihilate a one-particle state. This hypothesis permits us to obtain a field operator algebra similar to the creation and annihilation operator of the simple harmonic operators algebra. The Hilbert space with this construction is usually called the Fock space.  Therefore, Fock space is constructed with the tensorial product of one-particle Hilbert state \cite{tagahu}:
\begin{equation} \label{hfss} \mathcal{H}\equiv \mathcal{F}(\mathcal{H}^{(1)})=\left\{ \C, \mathcal{H}^{(1)}, \mathcal{H}^{(2)},\cdots, \mathcal{H}^{(n)}, \cdots \right\} \equiv \bigoplus_{n=0}^\infty \mathcal{H}^{(n)}\equiv e^{\mathcal{H}^{(1)}} \,, \end{equation}
where $\C$ is vacuum state, $\mathcal{H}^{(1)}$ is one-particle states and $\mathcal{H}^{(n)}$ is n-particle states. The n-particle states are constructed by the tensor product of one-particle states (for bosons, a symmetry product, $ \mathcal{H}^{(2)}=S\mathcal{H}^{(1)}\otimes \mathcal{H}^{(1)}$ and for fermions anti-symmetric products, $ \mathcal{H}^{(2)}=A\mathcal{H}^{(1)}\otimes \mathcal{H}^{(1)}$). 
 The annihilation operator on the one-particle state results in the vacuum state. The creation operator on the vacuum state results in the one-particle state. Although the Hilbert space of a physical system is unique, the choice of the basis is not unique, similar to the choice of the coordinate system in classical theory. The norm of the field operator (or annihilation operator) is an arbitrary constant, which must be fixed by the auxiliary condition such as geodesic spectral condition, KMS condition, Hadamard condition, etc \cite{brmo96}. By fixing this arbitrariness, the vacuum state is defined. 
 
\begin{figure}
\centering
\includegraphics[width=6cm]{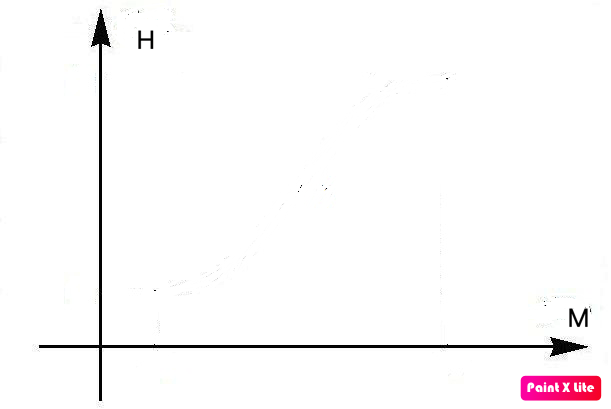}
\caption{Fock space bundle}
\label{fibu}
\end{figure}

In QFT, the field operator $\Phi(\vec x,t)$ plays a significant role in the connection between these two different spaces: a spacetime manifold $M$ and a Hilbert space $ \mathcal{H}$. On the one hand, it is immersed in spacetime, and on the other hand, it acts in Hilbert's space, which is defined at any point in a fixed classical spacetime background $M$ (of course in the distribution sense). Hilbert space can be thought of as the "fiber" of a bundle over the spacetime manifold, where each point of the manifold corresponds to a different fiber, $ \mathcal{H}\times M$ (figure \ref{fibu}). 
The bundle is typically referred to as a "Fock space bundle" \cite{taqg}. The Wightman two-point function  \cite{bailse,brmo96}, $\mathcal{W}(x,x')=\langle \Omega |\Phi(x)\Phi(x')|\Omega\rangle$, provides a correlation function between two different points in spacetime and their corresponding Hilbert spaces. $|\Omega\rangle$ is the vacuum state.

In quantum geometry \cite{taqg}, the quantum space of state is $ \mathcal{H}_{\mbox{total}} \equiv \mathcal{H}\times \mathcal{K}_{\mathcal{G}}$, where $\mathcal{K}_{\mathcal{G}}$ is the quantum space of states of space-time geometry. In the limit of the classical geometry, which is used in the quantum technology we obtain the Fock space bundle $ \mathcal{H}\times M$. $M$ can be considered as the laboratory frame and $ \mathcal{H}$ is the Hilbert space of the physical system in the laboratory.

\subsection{Quantum state}

In quantum theory, the physical system is modeled by a quantum state $\vert \alpha, t\rangle$ instead of particle $\vec x(t)$ and waves $\phi(t,\vec x)$. It is immersed and evolved in the Hilbert space. In the flat Minkowski space-time, the quantum state $\vert\alpha,t\rangle$ assume that it is the solution of the following field equation:
\begin{equation} \label{qmfe}
i\hbar \frac{\partial}{\partial t}\vert\alpha,t\rangle=H\vert\alpha,t\rangle,\;\; \mbox{or} \;\;\vert\alpha,t\rangle=U(t,t_0;H)\vert\alpha,t_0\rangle,
\end{equation}
where $H$ is the Hamiltonian of the physical system and $U(t,t_0;H)$ is the time evolution operator. To obtain the explicit form of the quantum state $\vert \alpha ,t\rangle$, we need to know the initial state $\vert \alpha ,t_0\rangle $. Although in quantum theory, the reality or results of the observation are probabilistic, if we have the initial state $\vert \alpha ,t_0\rangle $, the quantum state $\vert \alpha,t \rangle$ can be determined with certainty due to the unitary principle: $UU^\dag=\1$, where $\1$ is the identity operator on the Hilbert space. 

It is important to note that the trajectory of a particle $\vec{x}(t)$, the classical field $\Phi(\vec{x}, t)$, and the quantum state $|\alpha\rangle$ all of them are mathematical quantities used to explain physical systems. If our minds have difficulty grasping the quantum state, it is because the quantum state is a relatively new concept, whereas we are more accustomed to the concepts of a particle's trajectory and the field function. However, all these entities are abstract mathematical concepts, and within the microscopic domain, the quantum state provides a better explanation of our physical system than the other two quantities.

\subsection{Interaction and Observation}

In quantum theory, the act of observation or external interaction with the physical system causes the quantum state $\vert \alpha ,t\rangle$ to collapse into the new state $\vert \beta,t\rangle $. Evidently, it is not the same as the quantum state $\vert \alpha ,t\rangle$.  $\vert \beta,t \rangle $ is the reality of the physical system for the observer. If we call  $\vert \alpha ,t\rangle$ as the truth of the physical system, the reality of the physical system is different from the truth of the physical system. The reality depends on the observer and their chosen Hamiltonian detector $H_d$, but the truth depends on the boundary or initial conditions of the physical system and is independent of the observer \cite{takook2}. 

The Hilbert space of the interaction fields and the time evolution operator represent some of the most challenging problems in quantum technology. The time evolution operator cannot be defined properly because time is an observer-dependent quantity in the time-dependent interaction case. However, it can be defined approximately on a local level, making quantum technology and quantum computing theoretically feasible. One of the primary sources of error in quantum computation is the collapse of the quantum state during the interaction process, commonly addressed through Quantum Error Correction. This process must be defined in a manner independent of the observer. Another challenge in quantum computation is the high cost of isolating the quantum system, which leads to interactions with the environment, resulting in the collapse of the quantum state and subsequent errors in calculation.

A measurement of the physical system can be defined as a new system that includes the measuring apparatus. We do this by introducing the perturbed Hamiltonian $H_p=H_s+H_d$, where $H_s$ and $H_d$ are the Hamiltonians of the physical system and the detector, respectively. The observation of what we usually call particles (for example photons) or waves depends on the initial state $\vert\alpha,t_0\rangle$ and the Hamiltonian of the detector $H_{d}$ as well. If the initial state is one of the number operator's eigenstates, then the physical system's quantum state contains a definite number of particles or quanta. In this case, we can observe the photon if $[H_s,H_d]=0$. With a single photon, if $[H_s,H_d]\neq0$, the interference is observed or the wave properties of the photon. It is interesting to note that in the number states, the concept of particles or quanta exists but the number of the particles is dependent on the detector energy, which manifests the fluctuation in the Fock space.

If the initial state is a superposition of the particle number operator's eigenstates, the physical system's quantum state ($ \vert\alpha,t\rangle$) is a superposition of states, and the particle picture is inappropriate. One can only speak of the average number of photons or particles. We see that the particle-wave duality conundrum, well-known in ordinary quantum mechanics \cite{grza}, disappears in the framework of the quantum field theory as the fundamental concept in quantum field theory is the quantum state $\vert\alpha,t\rangle$. Expressing this concept in entangled states shatters classical physics' naive realism and locality. The reality observed by the observer depends on the apparatus of the observer, and reality observation is probabilistic, but the quantum state $|\alpha,t\rangle$ is deterministic due to the unitary principle ($UU^\dag=\1$).

In the general case, we must simulate the environment interaction by a Hamiltonian $H_e$ and define the total Hamiltonian as $H=H_s+H_d+H_e$. One of the challenges of implementing quantum logic gates is the delicate nature of quantum systems, which are highly sensitive to their environment. As a result, it is important to carefully control the noise and decoherence in quantum systems in order to implement reliable and accurate quantum logic gates.


\section{Quantum computation} \label{logicgate}

In quantum computation, classical bits are replaced by quantum state or qubits, which are represented by $\vert \alpha\rangle$ \cite{Nielsen}. Similarly, classical gates are substituted with quantum gates. These quantum gates act reversibly and are described by employing the unitary evolution operator $U$. Here, we discuss the fundamental concepts of quantum computation. The first important point in quantum computation is to build a deterministic and stable quantum state $\vert \alpha \rangle$. One of the technical challenges in quantum computation is isolating the physical system, or the qubits. Another challenge is the collapse of the quantum state during the registration and readout of a qubit. However, several techniques are available for reading out the state of a qubit in a deterministic manner. The realization of quantum technology comprises three steps: constructing an appropriate quantum state (registration), obtaining its time evolution (gate implementation), and detecting it deterministically (readout). 
\subsection{Quantum state construction}
Various methods exist for constructing a qubit or a quantum state such as trapped atoms and ions, photons in linear and non-linear optic, superconducting circuits, or quantum dots and etc. Among these, non-linear optics is discussed due to its ease of isolation \cite{meyer,slussa}. Non-linear optical processes can be utilized to generate single photons through the process known as spontaneous parametric down-conversion (SPDC). In SPDC, a pump laser beam is directed into a non-linear crystal, which splits the pump beam into two lower energy photons ($E=E_1+E_2$) that are entangled with each other. These entangled photons can be used as a resource for various quantum information tasks, such as quantum key distribution and quantum teleportation.  Another non-linear optical process that can be used for quantum information processing is second harmonic generation (SHG). In SHG, two photons with the same frequency are converted into a single photon with twice the frequency. This can be used to generate photons that are useful for quantum communication and computation. Increasing the number of qubits, creating unwanted entanglements between them makes it difficult to control the qubits.
\subsection{Quantum logic gate}
Quantum logic gates are a fundamental building block of quantum computing and perform unitary transformations on qubits. The basic quantum logic gates are the quantum NOT gate (also known as the Pauli X gate), the quantum Hadamard gate, and the quantum phase gate. The quantum NOT gate flips the state of a single qubit, while the Hadamard gate creates a superposition of two states. The phase gate introduces a phase shift into the state of a qubit.
There are also more complex quantum logic gates, such as the quantum controlled-NOT (CNOT) gate, which performs a NOT operation on a target qubit conditioned on the state of a control qubit. This gate is a key building block for many quantum algorithms, including quantum error correction and quantum teleportation.

To perform quantum logic gates, it is necessary to couple the qubits together so that they can interact with each other in a manageable way. The specific method used to couple the qubits can vary depending on the physical system being used but the general steps involved are similar. Once the qubits are coupled, a quantum logic gate can be applied to them by controlling the system's physical parameters. This includes the voltages applied to superconducting circuits, the laser beams used to manipulate trapped atoms and ions, and the polarization or number of photons in non-linear optics, etc. Finally, the output of the quantum logic gate must be measured in order to determine the result of the computation in a non-destructive way. For example, Raman transitions \cite{Grynberg} can be used to implement quantum gates, such as the Hadamard gate and the CNOT gate \cite{cizo}. 

\subsection{Register a quantum state}

Raman transitions are a fundamental technique in the field of laser spectroscopy and are used extensively in quantum optics and quantum information science. Raman transitions involve using two laser beams with different frequencies to drive a quantum system from one state to another. Specifically, the frequency difference between the two laser beams must match the energy difference between two states of the system, known as the "stimulated Raman resonance condition". When the two laser beams are focused on the quantum system, they create a spatially varying electric field, which causes the system to undergo a transition between the two states. This transition can be thought of as the absorption of one photon from one laser beam and the emission of another photon into the other laser beam, resulting in a transfer of energy between the two beams.

Raman transitions are used extensively in neutral atom quantum computing because they allow for the creation of superposition states, which are the basic building blocks of quantum information processing. By carefully tuning the frequency difference between the two laser beams, it is possible to create a superposition of the ground state and an excited state of an atom, which can then be used as a qubit for quantum information processing. The frequency difference between the two laser beams corresponds to the energy difference between the ground state and the excited state. The two laser beams are focused on the neutral atom, and the frequency difference is carefully chosen so that the energy difference between the ground state and the excited state is exactly matched. This causes the atom to undergo a Raman transition, which transfers the atom from the ground state to the excited state. After a short time delay, a second pair of laser beams is used to drive the atom back to the ground state, but with a slight phase shift that creates a superposition of the ground state and the excited state. This superposition state can be read out using the state-dependent fluorescence technique. 


 \subsection{Readout a quantum state} \label{readout}
 There are several techniques that can be used to read out the state of a qubit in a deterministic way, including projective measurement, weak measurement, quantum non-demolition (QND) measurement and dispersive readout \cite{dispersive}. These techniques can provide partial information about the state of a qubit without completely collapsing the state. The projective measurement on the qubit involves the application of a measurement operator that projects the qubit onto a specific state. For example, if we want to determine whether a qubit is in the state $|0\rangle $ or $|1\rangle$, we can apply a measurement operator that projects the qubit onto either the  $|0\rangle $ or $|1\rangle$ state. This will cause the qubit to collapse into the measured state with a probability equal to the squared magnitude of the corresponding coefficient in the superposition.
  
QND measurement involves measuring one observable in a way that does not affect the measurement of a second observable. The basic idea behind QND measurement is to use a measurement apparatus that is sensitive to the state of the quantum system but does not disturb the state of the system. For example, one common approach is to use an ancillary qubit that is coupled to the system of interest, and then perform a joint measurement on both qubits. By carefully engineering the interaction between the two qubits, it is possible to make a measurement of one observable of the system while leaving the other observable unchanged.


\section{Conclusion}

The main goal of this review is to present the fundamental concepts of quantum field theory in a simple way that will be useful for readers in the quantum computation domain. The relationship between qubits and logical gates with quantum states and unitary transformations are discussed. The various theoretical and technical challenges of quantum computation at a fundamental level are also recalled. One of the most challenging problems is the collapse of the quantum state during the manipulation of qubits, which forces us to read out the state of a qubit in a deterministic way.

\vspace{0.5cm}


\end{document}